\documentclass[a4paper,fleqn]{cas-sc}
\usepackage[authoryear,longnamesfirst]{natbib}

\def\tsc#1{\csdef{#1}{\textsc{\lowercase{#1}}\xspace}}
\tsc{WGM}
\tsc{QE}
\tsc{EP}
\tsc{PMS}
\tsc{BEC}
\tsc{DE}
\begin{document}
\let\WriteBookmarks\relax
\def\floatpagepagefraction{1}
\def\textpagefraction{.001}
\shorttitle{Using voice note-taking to promote learners' conceptual understanding}
\shortauthors{Khan et~al.}

%\begin{frontmatter}

\title [mode = title]{Using voice note-taking to promote learners' conceptual understanding}                      
\author[1]{Anam Ahmad Khan}
\cormark[1]
\ead{anamk@student.unimelb.edu.au}

\address[1]{School of Computing and Information Systems, The University of Melbourne, Melbourne, Australia}

\author[1]{Sadia Nawaz}
\ead{nawazs@student.unimelb.edu.au}

\author[1]{Joshua Newn}
\ead{joshua.newn@unimelb.edu.au}

\author[2]{Jason M. Lodge}
\ead{jason.lodge@uq.edu.au}
\address[2]{The University of Queensland, Brisbane, Australia}

\author[1]{James Bailey}
\ead{baileyj@unimelb.edu.au}

\author[1]{Eduardo Velloso}
\ead{eduardo.velloso@unimelb.edu.au}

\cortext[cor1]{Corresponding author}

\begin{abstract}
Though recent technological advances have enabled note-taking through different modalities (e.g., keyboard, digital ink, voice), there is still a lack of understanding of the effect of the modality choice on learning. In this paper, we compared two note-taking input modalities---\textit{keyboard} and \textit{voice}---to study their effects on participants' learning. We conducted a study with 60 participants in which they were asked to take notes using voice or keyboard on two independent digital text passages while also making a judgment about their performance on an upcoming test. We built mixed-effects models to examine the effect of the note-taking modality on learners’ text comprehension, the content of notes and their meta-comprehension judgement. Our findings suggest that taking notes using voice leads to a higher conceptual understanding of the text when compared to typing the notes. We also found that using voice also triggers generative processes that result in learners taking more elaborate and comprehensive notes. The findings of the study imply that note-taking tools designed for digital learning environments could incorporate voice as an input modality to promote effective note-taking and conceptual understanding of the text. 
\end{abstract}


\maketitle
\section{Introduction}
In a world where distractions abound, capturing learners' attention and encouraging deliberate reading is a significant challenge. Deliberate reading refers to the process of engaging with text to decode, understand and construct the meaning of the text \citep{afflerbach2008}. Such reading practices can enhance text comprehension and may support a better understanding of the reading material~\citep{ahmad2019}. One reading strategy that has been shown to be effective in supporting learners' text comprehension is to encourage them to take notes as they read~\citep{Liu2019, Nguyen2016, Ren2014}. Note-taking can help learners to plan learning activities and support the extraction of useful information from the learning material \citep{Witherby2019}. Previous research reports that note-taking can enable learners to understand and create knowledge---thereby increasing their learning performance ~\citep{kim2018,peverly2019, morehead2019a}. A likely reason for this is that note-taking may help learners' focus their attention, which in turn can help them encode the content into long-term memory~\citep{jansen2017}.

In recent times, there has been a shift in the note-taking medium from traditional longhand note-taking to more modern forms due to the increasing availability of portable electronic devices to take notes while reading (e.g.~keyboard \citep{morehead2019b} and digital ink \citep{kim2019}). Studies have found digital note-taking to be ubiquitous, faster, and beneficial for learners' memory \citep{luo2018, fiorella2017} and performance \citep{bui2013}. As compared to hand-written notes, digital notes offer some advantages such as they are easily searchable, editable, sharable and more malleable \citep{grahame2016}.

With the increased acceptance of voice as an input modality and advances in speech recognition technology, voice has emerged as a promising alternative note-taking modality. First, speaking has been found to consume fewer cognitive resources than writing thus permitting more attention to be directed to the text itself~\citep{Minyoung2018, grabowski2010}. Second, compared to writing, the act of speaking is faster and requires lower effort, thus allowing momentary thoughts to be recorded before they are forgotten \citep{shen2018}. Realising the potential of voice notes, a considerable number of note-taking tools such as \textit{Sononcent}\footnote{https://sonocent.com/}, \textit{Luminant}\footnote{https://luminantsoftware.com/apps/audionote-notepad-and-voice-recorder/}, and \textit{Notability}\footnote{https://www.gingerlabs.com/} have entered the market, enabling learners to not only record and index voice notes but also to access the voice recordings as transcripts at a later stage. The wide availability and adoption of these tools then pose questions regarding the impact of the note-taking modality on the learning experience. In particular, how does note-taking using voice changes learners' comprehension of text passages? Does a change in input modality from keyboard to voice influence learners' note-taking behaviours during digital reading? Are learners more meta-cognitively aware of their learning when taking voice notes? Answering these questions would not only help in understanding the impact of note-taking modalities on learners' conceptual understanding but would also inform the design of note-taking tools that can trigger higher-order learning while reading.  

Recent studies have investigated the effect of the note-taking medium on learners' performance \citep{luo2018, morehead2019b, fiorella2017}. These studies focused on comparing longhand pencil-paper to laptop note-taking. The findings of these studies suggest that learners record more information in their notes when typing them in the laptop as compared to the longhand writing \citep{fiorella2017}. However, note-taking using longhand can instigate a generative process that leads learners to take more elaborate notes \citep{luo2018}.  Either way, there is a clear effect of the note-taking medium on how learners' process and generate information, which have direct implications for effective note-taking. 

This led us to investigate whether switching the note-taking modality from typing to speaking would affect learners' note-taking behaviour, and if so, how would it impact learners' understanding of the digital text. We conducted an experiment in which 60 participants took notes using voice or keyboard on two independent pieces of digital text of comparable complexity. After the note-taking task, participants reported on their meta-comprehension level and performed a post-test, which included both factual and inference-based questions. We then built mixed-effects models to explore the effect of the note-taking modality on (1) learners' text comprehension, (2) the content of the notes and (3) learners' ability to make an accurate judgement regarding text comprehension. 

Our findings highlight the potential benefits of the use of voice for note-taking. Though participants taking notes with voice performed similarly to those taking notes with the keyboard on factual questions, they scored significantly higher on inference-based questions. This finding can be explained by the substantial differences in the content of the notes---voice notes tended to include more elaborations and more idea units than typed notes. However, this advantage was not reflected in participants' meta-comprehension judgement, as we found no significant effect of the note-taking modality on their ability to predict their learning on a future test. Our findings suggest that digital learning environments may incorporate voice as an input modality for effective note-taking, which can also assist learners in gaining a better conceptual understanding of the learning content. Given the fast speed and ubiquitous availability of voice input, our findings overall highlight a promising use case for voice notes in an educational context. 

\section{Related Work}
Studies have shown that a majority of learners take notes during learning activities \citep{peverly2019, morehead2019a}. From a cognitive psychology perspective, note-taking relates to information management where learners need to comprehend and write down the information that is personally meaningful to them \citep{piolat2005}. While taking notes, learners must filter the incoming information, update their existing knowledge, and integrate the newly processed information \citep{makany2009}. 

Research on note-taking suggests that the quantity of notes taken can vary substantially between learners and that the quality is difficult to account for \citep{peverly2019}. Some learners make a note of everything they hear in a lecture; others adopt a pick and choose a strategy where they take short and selective notes. The reasons for note-taking may also differ between learners, for example, some engage in note-taking as a ``process'' that can promote their recall and aid their concentration---known as the \textit{encoding effect}---while others take notes because of the resulting ``product'', which can help them in revising the material and keeping track of what was covered---known as the \textit{external storage effect} \citep{DiVesta1972}. Note-taking can also promote generative learning, as learners often relate the content they are reading to their prior knowledge \citep{morehead2019a}. The act of externalising their current knowledge can help learners realise meta-cognitive processes such as meta-comprehension (i.e.~learners' judgement of their level of comprehension \citep{dunlosky2007}), that can positively impact their decision to re-visit the material, and, in turn, can lead to improved learning~\citep{nguyenm2016}. 

Prior research has investigated the effects of note-taking on text comprehension \citep{moradi2020, ozccakmak2019, bahrami2017} and found that note-taking encourages learners to select, organise, and integrate ideas presented in the text. For instance, Bahrami et al.~investigated whether note-taking during reading improved learners' text comprehension. Their findings suggest that learners who take notes while reading were better able to integrate ideas between separate texts and performed better in post-reading tests than learners who did not take notes \citep{bahrami2017}.
 
Researchers have suggested that note-taking is not only a learner behaviour \citep{bauer2007} but also a study strategy \citep{karpicke2009}. Therefore, a better understanding of learners' note-taking is vital for both theoretical and practical reasons. While prior studies have been beneficial in understanding the role of note-taking on learners' conceptual understanding, they have mostly been focused on longhand note-taking using pencil and paper. As digital technologies become more prevalent in the classroom, learners increasingly take notes using portable digital devices, such as their laptops. Therefore, recent efforts have focused on investigating the effect of the medium (longhand vs.~laptop) on learners' note-taking \citep{mueller2014, fiorella2017, luo2018}. For instance, Mueller et al.~found that the medium of note-taking adopted by learners not only affects the content of notes generated but also their performance on a post-reading test \citep{mueller2014}. Their findings suggest that laptop note-takers recorded more words than longhand note-takers. However, longhand note-takers achieved higher performance on a post-reading test consisting of conceptual questions. Overall, the findings of studies investigating the impact of note-taking medium suggest that laptop note-taking can prompt transcription oriented or verbatim note-taking, which may not always boost learning performance \citep{fiorella2017, luo2018}. A likely reason for the poorer performance associated with verbatim notes is that they allow learners to record or note the material without any in-depth processing \citep{bauer2007}. 

With advances in multimedia sensing capabilities, researchers have been exploring other modalities beyond the keyboard for digital note-taking. For example, Han et al.~explored picture note-taking by presenting \textit{PicRemarkable}, a picture based note-taking tool which allows semantically related pictures with keywords to be included while taking notes \citep{han2014}. The findings of this study suggest that picture-based note-taking could support second-language readers to gain a better understanding of the reading material. Voice is another modality that has been gaining popularity. When compared to typing using the keyboard, voice can be faster to record and be more expressive \citep{yu2019}. Moreover, voice notes can also be transcribed to textual notes that, at a later stage, can make the notes review process easier. With the current development of voice note-taking tools and extensions for cloud-based document viewers such as \textit{Mote}\footnote{https://www.justmote.me/} and \textit{Kaizena}\footnote{ https://www.kaizena.com/}, learners can easily record voice notes. For example, \textit{Mote}, can assist learners in taking voice notes on digital documents and in labelling the voice recordings. Further, the recorded voice can automatically be transcribed to text and later, be shared with peers.  

Although tools supporting voice notes are starting to enter the market, there is a lack of research investigating the impact of voice note-taking on learning. Previous research suggests that the medium of note-taking not only can impact the content of the notes but may also affect learners' performance. In line with this, it logically follows that the note-taking modality adopted by learners can also influence their general note-taking behaviours, which, in turn, can potentially affect their learning performance. Therefore, in this paper, we aim to compare the two input modalities for note-taking in digital learning: \textit{keyboard} and \textit{voice}. We analyse the effects of these note-taking modalities on learners' text comprehension and meta-comprehension. Specifically, we aim to answer the following research questions: 
\begin{itemize}
\item RQ1: How does the note-taking modality affect learners' text comprehension? 
\item RQ2: How does the note-taking modality impact the contents of the notes made by learners? 
\item RQ3: Can note-taking modality influence learners' meta-comprehension judgement? 
\end{itemize}

\section{Experiment}
To investigate the effect of the modality of note-taking on learning, we conducted a within-subjects study in which learners practised note-taking with two different modalities while reading two passages digitally. Through the study, we analyse (1) the effects of modality on text comprehension, (2) the content of the notes and (3) the meta-comprehension judgement in a reading task.

\subsection{Participants}
We recruited 60 participants aged between 21--35 (M = 27.04,  SD = 4.39 ) via an online notice board from a large Australian university. 31 participants identified as men and 29 identified themselves as women. Participants were diverse in terms of their native language; with 16 English speakers, 3 Italian, 15 Mandarin, 4 Urdu, 5 Hindi, 7 Persian, 6 Sinhala, 2 Korean and 2 Tamil speakers. In terms of their education level: 29 participants were Masters students, and 31 were PhD students. As the reading task involved in the study related to the field of engineering, we ensured that none of the recruited participants were from an engineering discipline, reducing the potential familiarity with the subject matter. 

\subsection{Study Text}
We used two scientific text passages from a study by Mayer et al.~\citep{mayer1990}. These passages have been widely used in various studies for measuring the impact of note-taking and accurate meta-comprehension judgment on learners' performance \citep{nguyenb2016}. The first text passage comprised 792 words and was about different kinds of brakes and their functionality. The second passage discussed the mechanics of different types of pumps and had 850 words. We slightly modified the text in these passages so that both topics had a comparable reading difficulty level of 6.8, as measured by the Flesch-Kincaid Grade Level \citep{kincaid1975}. 

\subsection{Pre-test and post-test}
For each study text passage, we conducted a pre-test to test the participant's prior knowledge, and a post-test to test their text comprehension. The pre-test consists of six multiple-choice questions (MCQs). There were four candidate answers and one ``I don't know'' (IDK) option for each question. Following Jacob et al.'s approach, participants' text comprehension was assessed on a post-test that included \textit{factual} as well as \textit{inference}-based MCQs \citep{jacob2020}. For the post-test of each text passage, we developed 12 factual questions that required the knowledge of one idea unit presented in the passage (e.g., \textit{Light-weight bicycles use which kind of brakes?}). Further, for the post-test of each text passage, we developed six inference-based questions that required the integration of two or more idea units and tested the conceptual understanding of the text (e.g., \textit{A young lady in a village of India wants to pump water for cooking using a hand pump situated in her front yard. The water level in that area is 30 feet below the ground. What will happen when she moves the handle up and down?}). Similar to the pre-test, each post-test question had four candidate answers and one ``I don't know'' (IDK) option. All MCQs of the pre-test and post-test either received one point for the correct option or received zero points for incorrect or ``I don't know'' options.

\subsection{Meta-comprehension Judgement}
To investigate the impact of the note-taking modality on participants' meta-comprehension judgement, we asked them to report their judgement regarding their perceived performance on a knowledge test after each study phase. For this, after completing each reading task, participants were asked to answer the following question: \textit{Please indicate how confident you are that you can correctly answer questions about the read text (0 being no confidence, 100 being absolute confidence)}.

\subsection{Subjective Interest}
Note-taking behaviour may be impacted by learners' interest in the topic as well as their perceived difficulty~\citep{asher1978, barbier2006}. For this reason, we asked participants to rate their interest and perceived difficulty of the text on a 7-point scale from (1) ``\textit{Not at all interesting}'' to (7) ``\textit{Extremely interesting''}.

\subsection{Distraction Task}
We further included a distraction task in our study to prevent against the \textit{recency effect} of working memory \citep{ murdock1962}. The recency effect refers to learners ability to recall the most recently presented material better. To control for such an effect, we selected a distraction task that was unrelated to the research questions being investigated and was to be completed between each note-taking task and its corresponding post-test. Following Mundt et al., we asked participants to count backwards from 200 to 0 in multiples of seven for two minutes~\citep{ mundt2020}. In the second condition, participants performed the same distraction task but started counting from a different number (203). 

\subsection{Study Design and Setup}
To explore the impact of note-taking modality on participants' learning, we designed a within-subjects study with note-taking modality as a within-subjects factor (keyboard vs.~voice). Hence, all participants were exposed to both conditions of note-taking modality while reading a different text passage (i.e.~brakes or pumps) in each condition. The combinations of text passage and note-taking modality were counter-balanced across the sample. We facilitated the study entirely online through \textit{Zoom} video-conferencing application with \textit{Qualtrics}\footnote{https://www.qualtrics.com/} online survey tool alongside.

To collect the data for both conditions, we used \textit{Google Docs} with the \textit{Kaizena}\footnote{ https://www.kaizena.com/} add-on enabled as an online platform for participants to read text passages and to take notes using keyboard or voice. We opted for this particular tool because it is available inside \textit{Google Docs}, hence enabling participants to record voice notes by clicking on a recording button provided on the tool interface and without leaving the online text passage. Further, all participants also reported on their meta-comprehension judgment and answered the post-study test after each condition. Each study session was recorded with the consent of the participants and lasted approximately 90 minutes.

\subsection{Procedure}

\begin{figure}[t]
  \centering
  \includegraphics[width=0.8\linewidth]{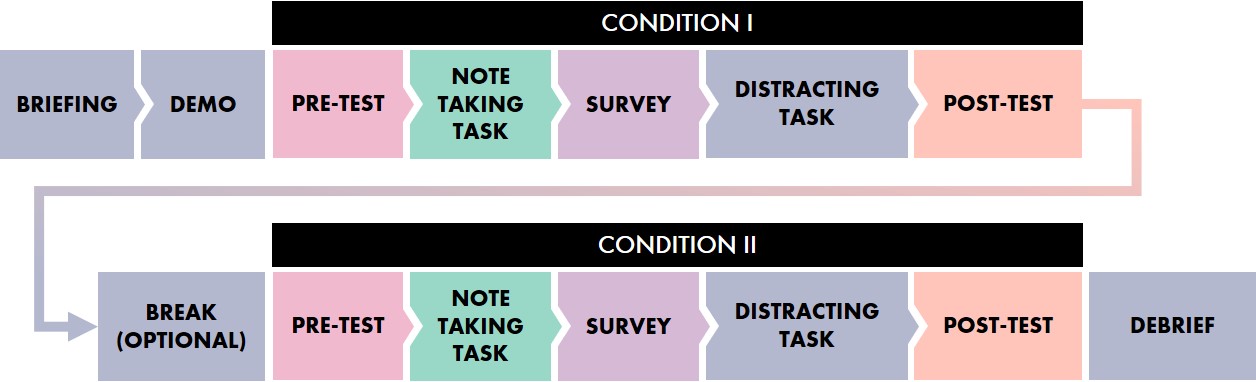}
  \caption{Study procedure adopted for exploring the effects of note-taking modality on participants' learning.}
   \label{fig:procedure}
\end{figure}

Figure \ref{fig:procedure} illustrates the study procedure. Once the participant connected to the session, the researcher gave a brief introduction to the study and the URL link to the Qualtrics page. The participant is then asked to read through the \textit{Plain Language Statement} and fill in the \textit{Consent Form} on the page. Upon obtaining the participant's written consent, the researcher shared their screen and demonstrated the note-taking task using Google Docs and Kaizena for both typed and voice notes. The researcher then instructed the participant to share their screen and to try note-taking for both modalities in the same environment. The participant will share their screen for the rest of the session henceforth.

For each condition, participants first completed a pre-test that consists of six MCQs. They then read the text passage and were asked to take notes that would help them in understanding the passage content and in answering the post-test questions using voice or keyboard, depending on the condition assigned. Before commencing the note-taking task, we informed participants that they would not be allowed to access or review the notes after the task or during the post-test. Participants were free to delete or re-record notes whenever they wanted, and there was no time limit imposed. The researcher stayed with the participant during the entire study session for them to respond to any questions or provide any technical assistance should they arise. However, during the note-taking task, both the researcher and the participant had their cameras and microphones turned off to minimise any experimenter bias. Immediately after completing the note-taking task, we asked the participants to complete a short survey segment (meta-comprehension and subjective opinion). Then, participants performed the distraction task for two minutes before answering the post-test. Between the two conditions, participants were given the option to take a short break (3 minutes maximum).

The order of the modality and the text passage was counterbalanced between participants, such that each participant took notes with both note-taking modalities but with different text passages. For instance, if in the first learning condition, a participant read the text passage on \textit{brakes} while taking notes using voice, then in the second learning condition, the same participant would read the text passage regarding \textit{pumps} while taking notes using the keyboard. At the end of the session, we debriefed participants and gave them a \$30 gift card for their time and contribution.

\section{Measures}
We investigate the effects of note-taking modality (voice vs.~keyboard) on participants' learning based on their text comprehension, the content of notes generated and their ability to make a judgement regarding the post-reading test.

\subsection{Text Comprehension}
We operationalised learners’ text comprehension in terms of their post-reading test score. The post-test consisted of 12 factual and 6 inference questions. The factual questions tested learners' ability to recall single idea units from the passage. In contrast, inference questions required participants to integrate multiple idea units across the text passage and tested their conceptual understanding of the passage.

\subsection{Note Content}
We analysed participants' notes based on the level of comprehension and elaboration. For the analysis of voice notes, we used a transcription service to manually transcribe the notes.

\subsubsection{Note Comprehensiveness}
Note comprehensiveness is measured by the \textit{number of idea units} stated in the notes. We considered idea unit as a concept consisting of an argument and its relations \citep{ kintsch1988}. For instance, the note: \textit{``Electric brakes are used as an electromagnet''} consists of one idea unit, in which \textit{``Electric brakes''} is an argument and \textit{``are used as an electromagnet''} is a relation. To measure note comprehensiveness, we first split each of the notes into individual statements and then counted the number of idea units. Two independent raters marked the number of idea units for 25\% of the notes. The overall inter-reliability rate as measured by Cohen's Kappa \citep{mchugh2012} was 0.95 -- suggesting a good agreement between the two raters. Hence, the remaining notes were coded by only one of the raters.

\subsubsection{Elaboration Level of Notes}
Elaboration Level is indicated by the \textit{number of elaborations} discussed in notes. Following Jacob et al., we considered an elaboration as an idea unit which was not explicitly stated in the text passage but was either an example or an analogy or participants' personal experience \citep{jacob2020}. For example, the following note contains two elaborations marked in italics as they were not stated in the text passage ``Seems like there are two kinds of pumps, one is the dynamic pump, and other is positive displacement pump. \textit{I would compare positive displacement to a dam}, and \textit{I will compare dynamic displacement to something driven by a motor}''. Two independent coders marked the number of elaborations for 25\% of the notes. The overall inter-reliability rate was 0.90, indicating good agreement between the two raters. Hence, only one of the raters marked the remaining notes.

\subsection{Meta-Comprehension}
We operationalised meta-comprehension judgement in terms of the meta-comprehension accuracy \citep{maki1998}, that is, the absolute difference between participants' meta-comprehension judgment and their performance on the post-test. A score of zero on this measure indicates an absolutely accurate judgment. The lower the value of this measure, the higher is the accuracy. The accuracy was calculated separately for each note-taking modality and for each participant.

\section{Results}
We first report the effects of note-taking modality on learners' text comprehension. Then, we report the analysis of note content when different note-taking modalities are used. Lastly, we investigate the effects of note-taking modality on learners' meta-comprehension judgement.

\subsection{Preliminary Analysis}
To better understand how the choice of text passages might have affected our results, we analysed the perceived difficulty and participant interest about them. We conducted separate paired t-tests for this analysis, and Cohen's \textit{d}~\citep{cohen2013} is reported as a measure of effect size. Table ~\ref{tab:preliminary} shows the mean perceived difficulty and interest across two text passages reported by participants. We did not find a significant difference in the perceived difficulty of the text passages, $t_{59}= -.77, p = .44, CI = [-.25,.5], d = .09$.  Similarly, we did not find a significant difference in participants' interest in the text passages, $t_{59} = 1.35,  p= .17, CI = [-.13,.70], d =  .17$. As such, we believe that the text passages that we had chosen were comparable.

\begin{table}
\caption{Mean with standard deviation (in parenthesis) of the perceived difficulty and interest across two text passages.}
\label{tab:preliminary}
\begin{tabular}{lcccl}
\toprule
Measure         & Brakes     & Pumps \\
\midrule
Difficulty (0-7) & 4.32 (1.76) & 4.04 (1.74)\\
Interest (0-7)   & 3.85 (1.64) & 4.23 (1.76)\\
\bottomrule
\end{tabular}
\end{table}

\subsection{Effect on text comprehension}
To investigate whether the note-taking modality affects participants' text comprehension, we built a linear mixed-effects model in \textit{R} using the \verb|lme4| package \citep{lme42015}. Linear mixed-effects models are used to represent dependent or clustered data which arise, for instance, when observations are collected over time for the same subject \citep{galecki2013}. It is an extension of the general linear model and includes both fixed effects and random effects. Variables that are expected to have an influence on the dependent variable are specified as \textit{fixed effects} in the model. In contrast, variables that are expected to have random influence on the dependent variable are specified as \textit{random effects}. The resulting model estimates the impact of the fixed effects on the dependent variable, after accounting for any influence of random effects. The rationale for using a linear mixed-effects model over a simple linear regression model was that the former can handle correlated data and unequal variances more effectively~\citep{wang2016}. As our design included several observations for the same participants, we built a mixed-effect model for the data to indicate both the random and fixed effects.

In this model, we treated the participant ID as a random effect to control for participants' idiosyncrasies. Participants in our study read a separate text passage in each of the two experimental tasks. To avoid any possible learning effects caused due to the differences between the text passage read in the two experimental conditions, we also included the passage ID as a random effect in the model. Participants' test scores were treated as the dependent variable. We treated the note-taking modality (Voice vs.~Keyboard), the question type (Factual vs.~Inference), and the interaction between note-taking modality and question type as fixed effects. Further, we added the participants' pre-test score for each passage as a covariate in the model.   

The final model, with its parameters, is shown in Table ~\ref{tab:lme}. To test for the assumptions of the linear mixed effect model, we visually inspected the residual plots, which did not reveal any apparent deviation from normality. Further, we checked for the presence of multi-collinearity between the fixed effects. All the predictors for the model had a variance inflation factor between 1 and 3, which is below the threshold to suggest multi-collinearity.

\begin{figure}[t]
  \centering
  \includegraphics[width=0.7\linewidth]{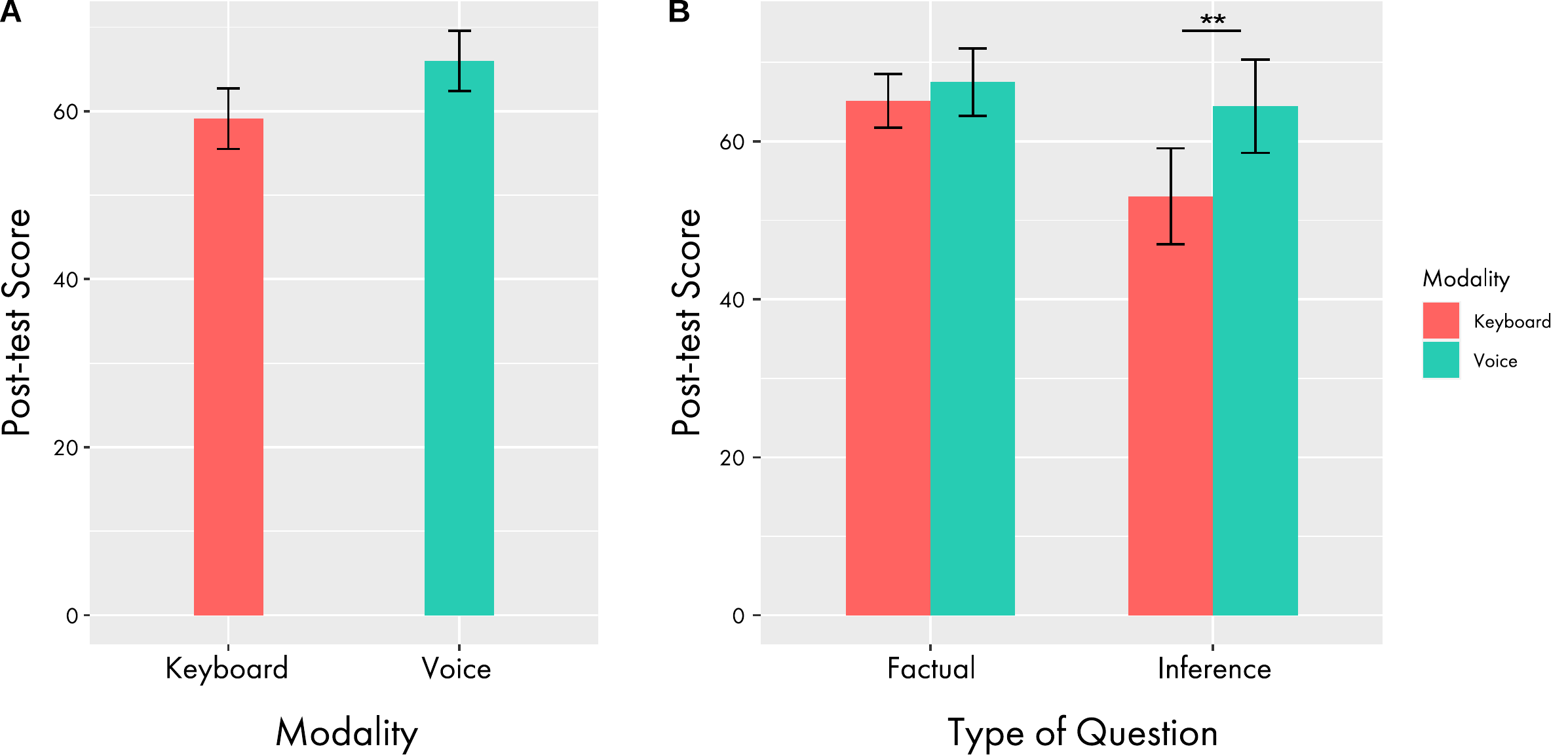}
  \caption{\textbf{A}: Overall Post-test scores across two conditions of note-taking modality. \textbf{B}: Post-test scores across two conditions of note-taking modality split on the question type. Error bars represent 95\% CI, **p < 0.05}
   \label{fig:mod}
\end{figure}

We measured the effect of note-taking modality on participants' text comprehension. We report this in terms of their overall post-test scores. The t-values and p-values of test results were reported on Satterthwaite's approximation of the effective degree of freedom \citep{satterthwaite1946}. Figure \ref{fig:mod} presents a bar chart of participants' post-test scores across the two note-taking modalities. 

The model is statistically significant ($\chi^2_4 = 27.29, p < 0.01$) and describes 9\% of variance of participants’ test scores (Marginal $R^2 = 0.09$, Conditional $R^2 = 0.29$). The results indicate that although the mean post-test score was higher for voice note-taking (mean = 65.2\%, SD = 19.9) than keyboard note-taking (mean = 59.1\%, SD = 19.1), the effect of modality in predicting learners' score was not significant across all questions ($p > 0.05$). 
\begin{table}
  \caption{Effects of model factors on predicting participant’s test score. Model Formula = $Score \sim Modality * QuestionType + PriorKnowledge + (1|Participant) + (1|Passage)$, where $Score$ = post-test score, $Modality$ = note-taking modality (Voice vs.~Keyboard), $QuestionType$ = type of question (Factual vs.~Inference), $PriorKnowledge$ = pre-test score, $Participant$ = identification number of each participant and $Passage$ =  identification number of each passage, ***p < 0.001, *p < 0.05 }
  \label{tab:lme}
  \begin{tabular}{lcccl}
    \toprule
    Variable&Estimate&SE&$t$&$p$\\
    \midrule
    (Intercept)&16.52&4.3&5.15&2e-16***\\
    Modality (Keyboard) &-2.46&3.09&-0.79&0.42\\
    QuestionType (Inference) &-3.05&3.09&-0.98& 0.32 \\
    PriorKnowledge &-0.18&0.10&-1.7& 0.2 \\
    Modality:QuestionType&-3.40&4.01&-2.40&0.04*\\
  \bottomrule
\end{tabular}
\end{table}

\begin{table}[pos=b]
  \caption{Pairwise contrast of interaction between note-taking modality and question type, **p < 0.01, ***p < 0.001}
  \label{tab:pairwise}
  \begin{tabular}{lcccccl}
    \toprule
    Contrast&Estimate&SE&$t$&lower CI &upper CI& $p$\\
    \midrule
    Voice Factual - Keyboard Factual&2.46&3.12&0.78&-5.63&10.56&0.85\\
    Voice Factual - Voice Inference&3.05&3.12&0.98&-5.04&11.15&0.76\\
    Keyboard Factual - Keyboard Inference&12.08&3.12&3.87&3.99&20.18&0.000*** \\
    Voice Inference - Keyboard Inference&11.49&3.12&3.68&3.40&19.59&0.001**\\
  \bottomrule
\end{tabular}
\end{table}
As expected, participants' scores in the inference-based questions (mean = 58.8\%, SD= 23.8) were lower than the factual questions (mean = 66.31\%, SD = 14.9). However, this difference was not statistically significant ($p > 0.05$). 
Finally, we observed a significant interaction effect between modality and question type ($t_{178} =-2.40, p = 0.04$). This result suggests that the note-taking modality has different effects depending on the type of question being asked. To explore the interaction between the modality and question type, we conducted pairwise comparisons from the contrasts between factors using the \verb|lsmeans| package in R \citep{lsmeans2016}. Table \ref{tab:pairwise} shows the results of differences in the least square (LS) means with the associated confidence intervals (CI) and p-values for pairwise comparison between the two factors. 

The results in Table \ref{tab:pairwise} and Figure \ref{fig:mod} highlight two interesting insights. First, whereas we found no significant effect of the note-taking modality on test scores in factual questions, ($CI = [-5.6,10.5], p>.05$), we did find a significant effect in inference-based questions ($CI = [3.4,19.6], p < .01$). Second, whereas we did not observe a significant difference between factual and inference-based question scores when using voice notes ($CI = [-5.0, 11.1], p=.76$), we found a significant drop in performance in inference-based questions when using typed notes ($CI = [4.0,20.1], p<.01$).   

\begin{figure}[b]
  \centering
  \includegraphics[width=1\linewidth]{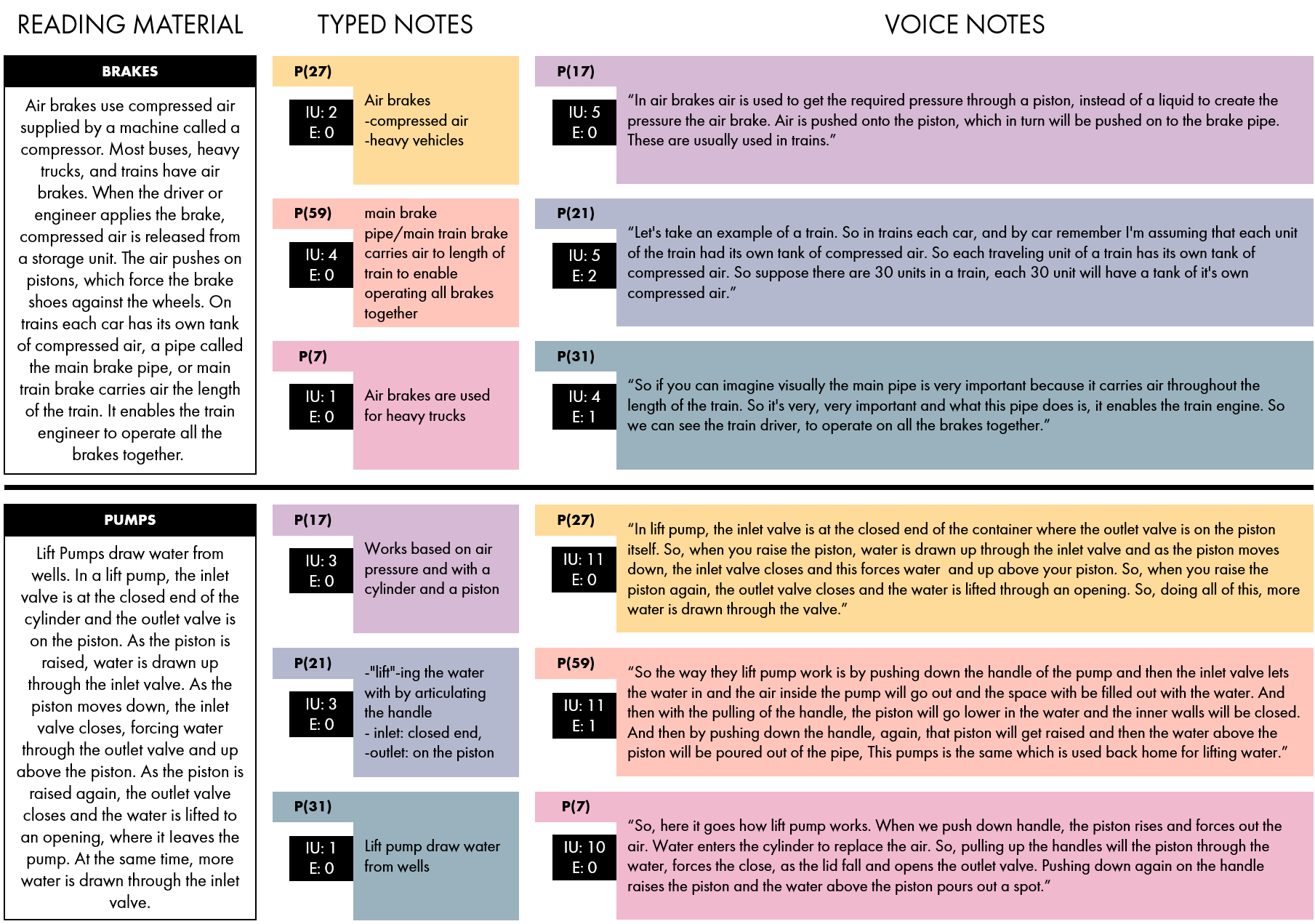}
  \caption{Typed and voice notes made by participants (ID given in parenthesis) for a paragraph of the two text passages. \textbf{IU} indicates the number of idea units discussed in each note. \textbf{E} indicates the number of elaborations made in each note }
   \label{fig:notesmodality}
\end{figure}

\subsection{Effect on note content}
Given that the affordances of the input modality can shape user behaviour, we expected to see differences in the content of the notes depending on the choice of modality. To explore the effect of the modality on the note content, we analysed the notes in terms of their comprehensiveness and and level of elaboration. Comprehensiveness reflects the number of idea units discussed in notes, while the level of elaboration refers to the number of elaborations of an example, analogy and personal experience discussed in notes. We built two separate linear mixed-effects models to investigate this effect. In both models, to control for the idiosyncrasies of the participants and the text passages, we treated the participants and passage IDs as random effects. We treated the note-taking modality (voice vs.~keyboard) as a fixed effect for both models. The number of elaborations and the number of idea units were treated as the dependent variables. Figure \ref{fig:notesmodality} shows representative examples of the kinds of notes made by participants by using the two note-taking modalities.

\begin{figure}
  \centering
  \includegraphics[width=0.7\linewidth]{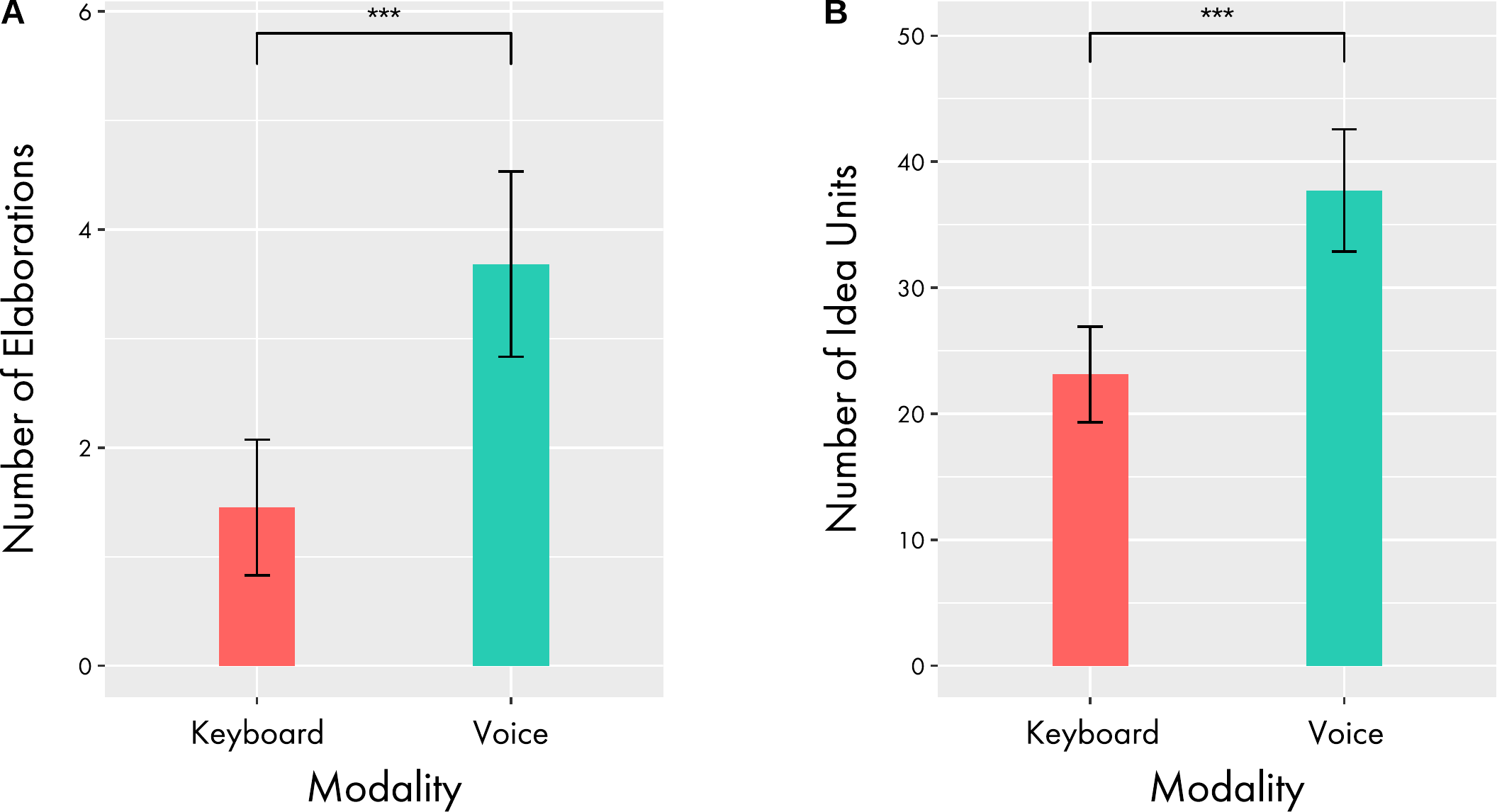}
  \caption{\textbf{A}: Total number of elaborations made by participants using voice and keyboard note-taking modality. \textbf{B}: Total number of idea units discussed by participants using voice and text note-taking modality. Error bars represent 95\% CI, ***p < 0.001 }
   \label{fig:ideaUE}
\end{figure}

The model was statistically significant ($\chi^2_1 =24.9, p < .001$) and described 13\% of variance (marginal $R^2 = .13$, conditional $R^2 = .45$). The main effect of modality on note comprehensiveness was significant ($d =  1.63, p < .001$).  The model is shown in Table \ref{tab:combined}. The model suggests that when taking voice notes, learners include more idea units than when typing notes by using a keyboard (see Figure \ref{fig:ideaUE}).

\begin{table}
  \caption{Effect of modality on predicting note comprehensiveness, indicated by number of idea units and level of elaboration indicated by the number of elaborations, **p< 0.01, ***p < 0.001 }
  \label{tab:combined}
  \begin{tabular}{lccccl}
    \toprule
     &&Note Comprehensiveness&&\\
    \midrule
    Variable&Estimate&SE&$t$&$p$\\
    \midrule
    Intercept&23.76&3.0&7.89&2.84e-12***\\
    Modality&18.6&3.38&5.5&0.000***\\
    \midrule
    &&Elaboration level of notes&& \\
    \midrule
    Intercept&1.45&0.46&3.14&0.002**\\
    Modality&2.53&0.64&3.93&0.000***\\
    
  \bottomrule
\end{tabular}
\end{table}

The model for the level of elaboration was also statistically significant ($\chi^2_1 = 14.28, p < 0.001$) and described 11\% of variance (marginal $R^2 = .11$, conditional $R^2 = .13$). The main effect of modality on elaborations was significant ($d = 0.69, p < .001$). The model is shown in Table \ref{tab:combined}. These results suggest that learners tend to elaborate more on the content when taking voice notes than when typing notes (see Figure \ref{fig:ideaUE}).

\begin{figure}[b]
  \centering
  \includegraphics[width=0.6\linewidth]{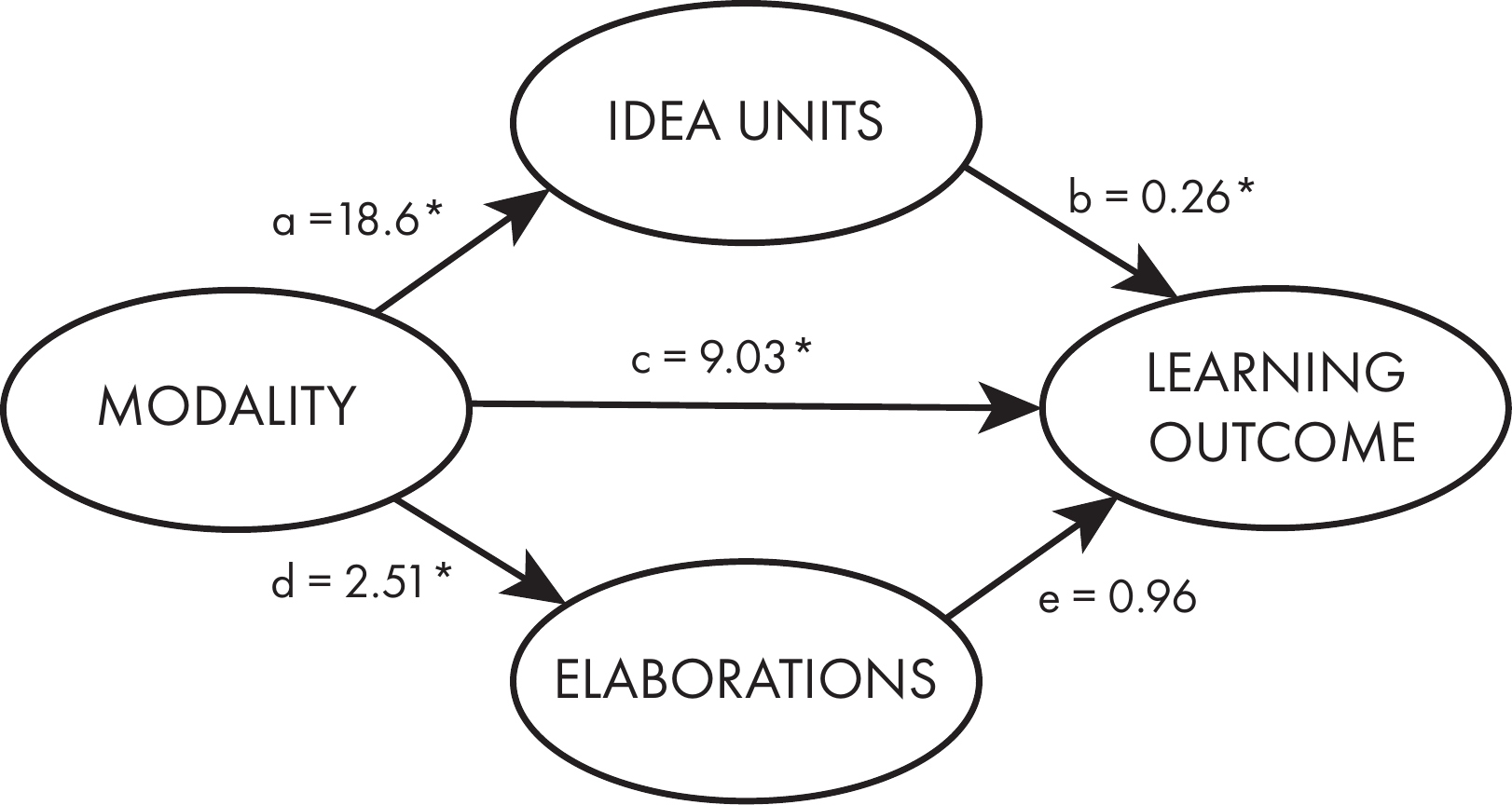}
  \caption{The model showing the effects of note taking modality on learners' learning performance on inference based question, with the number of idea units and number of elaborations as mediator. Values on paths are regression coefficients. * p < 0.05}
  \label{fig:mediation}
\end{figure}

These findings suggest a potential explanation for the positive effect of the use of voice notes on participants' performance on inference-based questions---because voice notes lead learners to increased elaboration and comprehensiveness in their notes by discussing more idea units together, they lead to improved learning. To test this hypothesis, we conducted a mediation analysis by taking the number of elaborations and the number of idea units as mediators. Note-taking modality was the dummy-coded predictor variable, and learners' performance on the post-test inference-based questions was taken as a dependent variable. To conduct the mediation analysis, we applied Hayes's bootstrapping methodology \citep{hayes2017} with 10,000 simulations. The result of the mediation analysis is shown in Figure \ref{fig:mediation}. We report the results of the significant effect of the mediation analysis using the estimates, SE, and 95\% CI of the estimates. The CI estimates that do not include zero are considered significant. 

The findings of the mediation analysis revealed a significant direct effect of the note-taking modality on the learning performance (estimate = 9.03, 95\% CI [.05, 18.01]). Additionally, we found that modality had a significant indirect effect of 4.77, SE= 2.21, 95\% CI [.82, 9.95], on learners' learning performance mediated via the number of idea units. On the other hand, regarding the number of elaborations, the analysis revealed a non-significant indirect effect of 2.41, SE= 1.75, 95\% CI [-6.06, 0.88] of modality on learners' learning outcome. The results of the mediation analysis suggest that voice leads learners to elaborate more on the information read in the text passage and to make more comprehensive notes, indicated by the number of idea units.  However, only the note comprehensiveness mediated the effect of note-taking modality on learning performance for inference-based questions.

\subsection{Effect on meta-comprehension accuracy}
The meta-comprehension accuracy was operationalised as the absolute deviation between learners' judgement and their actual test performance on the entire post-test. The mean meta-comprehension for the voice note-taking condition was 17.25 (SD = 11.93), and for keyboard note-taking condition was 15.09 (SD=11.9). To measure the effect of note-taking modality on learner's meta-comprehension judgement, we again built a linear mixed effect with the absolute meta-comprehension accuracy as a predictive variable, the participant and passage IDs as random effects, and the note-taking modality as a fixed effect. The model was not statistically significant ($\chi^2_1 = 1.07, p =0.29$) and the main effect of modality on learner's meta-comprehension judgement was not significant ($t_{60} = -1.43, p = .30$). The results indicate that even though the mean meta-comprehension accuracy reported by learners using keyboard as a note-taking modality was higher than voice, we found no significant effect of the note-taking modality on participants' ability to predict their performance on a future test.

\section{Discussion}
In this paper, we investigated whether note-taking modality could affect learners' text comprehension, their note contents, and their ability to make an accurate judgement regarding text comprehension. By analysing these behaviours, we aimed to develop a better understanding of voice note-taking modality on learning. In this section, we discuss the findings of our research questions and some implications of our work.

\subsection{Towards a better conceptual understanding of text}
Our first research question examined how note-taking modality can affect learners' text comprehension. To measure this, we assessed learners' performance on the post-test, which consisted of factual and inference-based questions. Our findings suggest that typed notes and voice notes may offer similar benefits when answering factual questions. These findings are in line with prior research that investigated the effect of explanatory modality on learners' text comprehension  \citep{jacob2020, lachner2018}. \citet{jacob2020}, for instance, found that if the goal of the learning activity is to acquire basic knowledge (as opposed to deeper knowledge), then the modality used for articulating the explanation may not affect participants' performance. 

However, we found that voice notes seemed to offer an advantage in answering inference-based questions. A possible explanation for this may be that inference-based questions required learners to connect multiple idea units across the text passage and tested their conceptual understanding of the passage. Based on learners' notes, we found that the number of idea units in voice notes (mean idea units/voice note = 5.14 $ \pm $ 3.7) far exceeded the number of idea units in typed notes (mean idea units/text note =  2.36 $ \pm $ 1.93). Thus, it appears that voice note-taking may assist learners in generating inferences based on the information presented in the text passages, where learners seemed to connect the ideas and concepts across those passages. The act of integrating idea units across multiple sentences may lead to the construction of new knowledge \citep{eglington2018, clinton2014}. Learners who are able to discuss more idea units in a single note are likely to make better inferences about the text, and are also likely to develop a better conceptual understanding, which is reflected in their performance in the post-test. These findings were also in line with prior studies that investigated the effect of explanatory modality on inference-based questions~\citep{jacob2020, lachner2018}, whose results suggest that while generating explanations using voice, learners elaborated more on the text passage, which increased their score in inference based questions~\citep{jacob2020}.

We also found that learners’ performance was significantly lower for inference-based questions than the factual questions when using keyboard---which is the dominant note-taking modality in most current digital note-taking applications. We believe that a possible reason for this could be that generally, learners used fewer idea units in their notes (see Figure \ref{fig:ideaUE}). Moreover, the number of elaborations in typed notes were either very few or none at all (see Figure \ref{fig:ideaUE}). Therefore, after typing notes using a keyboard, while learners could recall specific idea units in factual questions, they struggled to connect multiple idea units when they were required to answer inference-based questions.

\subsection{Towards generation of better notes}
Our second research question examined the impact of note-taking modality on the content on notes generated by learners. We analysed the contents of notes in terms of their comprehensiveness and level of elaboration. We defined comprehensiveness in terms of the number of idea units discussed in a note. Further, we measured the level of elaboration in terms of the number of elaborations made by the learner. We found that learners who used voice as a modality for their notes discussed more idea units than those who used keyboard as a note-taking modality. Similarly, learners elaborated more on the subject matter when they made notes using voice modality than when they used the keyboard modality. A possible reason for this behaviour could be the presence of social involvement \citep{ fiorella2019}. As suggested in prior research \citep{ jacob2020, lachner2018}, voice can encourage learners to address a potential audience by making more person-deictic references (e.g., ``me'', ``you'') in their notes \citep{ lakoff1982, sindoni2014}. This higher levels of social involvement could potentially lead learners to discuss more idea units and elaborate more on the subject matter \citep{ fiorella2019}. 

Further, we observed that note comprehensiveness (indicated by the number of idea units) mediated the effect of note-taking modality on text comprehension. These findings suggest that voice notes led learners to take more comprehensive notes, which, in turn, improved inferences about the text, which ultimately resulted in a higher score for inference-based questions. 

Based on previous research \citep{fiorella2019, lachner2018}, we expected that the number of elaborations could also mediate the effect of modality on learners' text comprehension. However, our mediation analysis findings did not reveal such an effect. We suggest two possible explanations for this finding. First, in previous studies, participants were explicitly asked to generate explanations; the act itself can lead them to provide examples by relating the material to their prior knowledge \citep{fiorella2016}. In contrast, we asked participants to practice note-taking without explicitly asking them to elaborate on the content. It is quite possible that some participants took notes to summarise the content without elaborating on it. Second, participants in our study had no prior experience with the subject matter; which made it more difficult to relate the content of the text to real-life examples. As a result, the total number of elaborations in notes was not large enough to mediate a significant effect of modality on text comprehension (see Figure \ref{fig:ideaUE}).

\subsection{Comparable effects of modality on meta-comprehension}
The last research question investigated the effect of note-taking modality on learners' meta-comprehension judgement. The results of our analysis revealed no significant effect of note-taking modality on learners' ability to judge their performance for an upcoming test. From the results of this study, we suggest that even though a change in note-taking modality from keyboard to voice increases learners conceptual understanding of the text, the perception of their learning performance might remain unaffected. These results are consistent with the findings of prior research \citep{fukaya2013, maki2005}, which revealed no significant correlation between learning performance of participants on immediate post-reading test and their meta-comprehension judgement. The findings of these studies suggested that even if learners engage in active elaborations which leads them to better comprehend the text and to provide comprehension-level cues, learners may not use these cues when judging their own text comprehension.  

\subsection{Implications}
We believe that the findings of the study have educational implications. With the pervasiveness of digital learning environments where learners are provided with digital content, incorporation of voice as an input modality for note-taking activity can support learning goals. Compared to typing notes using a keyboard, voice note-taking can encourage learners to reflect and ponder on their understanding of the subject matter, thus assisting them in developing a better conceptual understanding. Moreover, voice note-taking may also allow learners to focus more on the content itself rather than being distracted by typing errors or formatting issues which can cause distraction when notes are being typed. Thus, designing note-taking tools for the digital learning environment, which includes voice as an input stream for recording notes, and encouraging learners to practice note-taking using voice may help them to gain a better conceptual understanding of the content. 

From instructors' view-point, incorporating voice as an input modality for note-taking can also create possibilities for them to provide verbal feedback to learners. The feedback provided in the form of voice notes by instructors can be more meaningful and better understood by learners, as voice has intrinsic features of prosody that can communicate a significant amount of semantic and syntactic information \citep{akker2003, cutler1997}.

\section{Limitations and Future Work}
We acknowledge four limitations in this study. Previous research indicates that the effects of modality used for learning task on text comprehension depend on the complexity of the text passage used in the learning task. For instance, Jacob et al.~suggest when generating explanations, the effect of modality on text comprehension depends on the text complexity~\cite{jacob2020}. Whereas the effect of the modality is small in texts of lower complexity, the effect is more pronounced in texts of higher complexity. In the present study, the reading material used for note-taking was perceived to have high levels of text complexity by the novice learners (mean subjective reading difficulty reported = 4.18,  SD = 1.76). Hence our findings regarding the superior effects of voice notes for the conceptual understanding of the reading material are suitable for text passages that are perceived to have a higher level of text complexity. Future research should investigate whether these findings could be generalized with reading materials with a lower level of text complexity.

Further, to observe the effects of note-taking modality on learners' text comprehension, we conducted the post-test immediately after reading and distraction task. As a consequence, we only examined the short-term effects of the note-taking modality on text comprehension. The question of whether these effects translate to a longer-term remains open for future work to explore, possibly by conducting a delayed knowledge test.

A third limitation is that in this study participants were not allowed to review their notes before the post-reading test. The rationale for this design decision was that reviewing typed and voice notes could demand varying time which could potentially be a factor in the study. Nevertheless, future research could be conducted to allow learners to review their notes during the reading task to investigate the impact of the review of voice notes on text comprehension. In particular, we note the distinction between the input modality used to create the note and the output modality used to consult it. Through the use of text-to-speech and speech-to-text technologies, it is feasible to separate the creation and consumption of notes. As such, future work can explore the different combinations of typed/spoken notes as input modalities and written/audio notes as output modalities. 

Lastly, we observed that while taking voice notes, learners recorded information that did not require spatial structuring. However, many areas of knowledge can benefit from a visual representation, which includes anything from equations to mind maps. In these cases, the voice modality may not be as effective for learning as compared to other modalities. Therefore, our findings regarding the superior effects of voice notes on learning are limited to the reading content regarding which learners can easily record the information using the voice modality without needing to draw diagram or concepts maps in their notes. This suggests that note-taking applications should provide users with an alternative or complementary input modalities, enabling users to decide how to best leverage the affordances of each of them.

\section{Conclusion}
Although voice note-taking is becoming increasingly popular, there is a lack of evidence about how this modality may support learning goals. The present study bridges this gap by providing insights into the effects of voice note-taking on learners' text comprehension, their meta-comprehension judgements, and the contents of their notes. Our findings suggest that a change of modality from keyboard to voice may enable learners to make more elaborations and discuss more idea units in their notes. This can assist learners in text comprehension, as reflected in their post-test scores. Further, the findings indicate that if the learning goal is to gain basic knowledge of the study content, then both modalities (keyboard vs.~voice) perform similarly, but if the goal is to enable learners to gain a conceptual understanding by making inferences, voice note-taking might be more effective than typing notes using keyboard. Lastly, the findings suggest that both note-taking modalities have comparable effects on learners' meta-comprehension judgement of text comprehension. Overall, our findings suggest that voice modality could be incorporated in digital learning environments for taking notes as it could trigger higher-order learning and enable learners to get a better conceptual understanding of digital texts.

\appendix
\printcredits
\bibliographystyle{cas-model2-names}
\bibliography{cas-refs}

\end{document}